\documentclass[twocolumn,aps,prc,floatfix]{revtex4}
\usepackage{graphicx}
\usepackage{dcolumn}
\usepackage{bm}
\usepackage{mhchem}

\begin{document}

\title{Modeling pion production in heavy-ion collisions at intermediate energies}

\author{Gao-Chan Yong}%\email{yonggaochan@impcas.ac.cn}
\affiliation{%
Institute of Modern Physics,
Chinese Academy of Sciences, Lanzhou 730000, China
}%

\begin{abstract}
Pion production in nucleus-nucleus collisions at intermediate energies was modeled in the framework of the Isospin-dependent Boltzmann-Uehling-Uhlenbeck (IBUU) transport model. The effects of nucleon-nucleon short-range correlations in initialization and mean-field potential, isospin-dependent in-medium baryon-baryon elastic and inelastic cross sections and pion in-medium effect are all considered in this model. It is found that the ratio and yields of $\pi^{-}$and $\pi^{+}$ in Au+Au reaction at 400 MeV/nucleon reproduce the FOPI/GSI data very well especially with a soft symmetry energy in the present transport model. Predictions on the single and double $\pi^{-}/\pi^{+}$ ratio are made for the isotope reaction systems $^{132}\rm {Sn}+^{124}\rm {Sn}$ and $^{108}\rm {Sn}+^{112}\rm {Sn}$ at 300 MeV/nucleon since related experiments are being carried out at RIKEN/Japan.

\end{abstract}

%\pacs{25.70.-z, 21.65.Mn, 21.65.Ef}

\maketitle

\section{Introduction}

The equation of state (EoS) of
nuclear matter at density $\rho$ and isospin asymmetry
$\delta$ ($\delta=(\rho_n-\rho_p)/(\rho_n+\rho_p)$) can be
expressed as \cite{li08,bar05}
\begin{equation}
E(\rho ,\delta )=E(\rho ,0)+E_{\text{sym}}(\rho )\delta ^{2}+\mathcal{O}%
(\delta ^{4}),
\end{equation}%
where $E_{\text{sym}}(\rho)$ is nuclear symmetry energy.
Nowadays the EoS
of isospin symmetric nuclear matter $E(\rho, 0)$ is relatively well
determined \cite{pawl2002} but the EoS of isospin
asymmetric nuclear matter, especially the high-density behavior
of the nuclear symmetry energy, is still very uncertain \cite{Guo14}.
There are plenty of studies showing inconsistent
results on pion production \cite{wolter06,xie13,xiao09,prassa07,feng10,hong2014,Reisdorf07,cozma17} when comparing theoretical simulations to the data from FOPI detector \cite{Reisdorf07} at GSI.
Constraints on the high-density behavior of the symmetry energy from ground-based measurements can be highly relevant to neutron stars \cite{Lat01}, such as their stellar radii and moments of inertia, crustal vibration frequencies and neutron star cooling rates \cite{Lat04,Vil04,Ste05}.
Experimentally, related measurements of pion and nucleon, triton and $^{3}$He yields ratio in the isotope reaction systems $^{132}\rm {Sn}+^{124}\rm {Sn}$ and $^{108}\rm {Sn}+^{112}\rm {Sn}$ at 300 MeV/nucleon, are being carried out at RIBF-RIKEN in Japan \cite{sep,shan15}.

The mechanism relating to pion production in heavy-ion collisions at intermediate energies
has increasing interests recently \cite{ko1,ko2,ko3,ko4,pi3,cozma16,cozma17}.
To demonstrate pion production in heavy-ion collisions at intermediate energies,
I use the newly updated Isospin-dependent
Boltzmann-Uehling-Uhlenbeck (IBUU) transport model. This model includes the
physical considerations of nucleon-nucleon short-range correlations, the isospin-dependent in-medium elastic and inelastic baryon-baryon cross sections as well as the momentum-dependent isoscalar and isovector pion potentials \cite{yong20151,yong20152,yong20153}. With a soft symmetry energy in the newly updated transport model, the ratio of yields of $\pi^{-}$and $\pi^{+}$ in Au+Au reaction at 400 MeV/nucleon fits the FOPI/GSI data very well. To further compare with experimental data so as to obtain the information of the density-dependent symmetry energy, the single and double $\pi^{-}/\pi^{+}$ ratios are predicted for the isotope reaction systems $^{132}\rm {Sn}+^{124}\rm {Sn}$ and $^{108}\rm {Sn}+^{112}\rm {Sn}$ at 300 MeV/nucleon since related experiments are being carried out at RIKEN/Japan.

\section{Modeling pion production in the IBUU transport model}

The present Isospin-dependent
Boltzmann-Uehling-Uhlenbeck (IBUU) transport model has its origin from the IBUU04 model \cite{lyz05}.
The BUU transport model describes the time
evolution of the single particle phase space distribution function
$f(\vec{r},\vec{p},t)$, which is given by
\begin{equation}
\frac{\partial f}{\partial
t}+\nabla_{\vec{p}}E\cdot\nabla_{\vec{r}}f-\nabla_{\vec{r}}E\cdot\nabla_{\vec{p}}f=I_{c}.
\label{IBUU}
\end{equation}
The phase space distribution function $f(\vec{r},\vec{p},t)$
denotes the probability of finding a particle at time $t$ with
momentum $\vec{p}$ at position $\vec{r}$. The left-hand side of
Eq.~(\ref{IBUU}) denotes the time evolution of the particle phase
space distribution function due to its transport and mean field,
and the right-hand side collision item $I_{c}$ accounts for the
modification of the phase space distribution function by elastic and
inelastic two body collisions.
$E$ is a particle's total energy, which is equal to
kinetic energy $E_{kin}$ plus its average potential energy $U$.
The mean-field potential $U$ of the single particle depends
on its position and momentum of the particle and is given
self-consistently by its phase space distribution function $f(\vec{r},\vec{p},t)$.

In the updated model, the initial density distributions of neutron and proton in nucleus are given by the Skyrme-Hartree-Fock with Skyrme
M$^{\ast}$ force parameters \cite{skyrme86}. The proton and neutron
momentum distributions with high-momentum tail reaching about twice local Fermi momentum are used \cite{yong20151,yong20152}. Experiments show that for medium and heavy nuclei there is a rough 20\% depletion of nucleons (caused by the neutron-proton correlations) with momenta distributed above the Fermi momentum \cite{sci08,sci14}.
The depletion effect is caused by dynamical self-energies due to
short range correlations and core polarization. However,
the present microscopic theoretical calculations based on the nucleon self-energies
can not quantitatively reproduce these experimental findings, some phenomenological methods
are thus used.
I let nucleon momentum distribution in the high-momentum tail
\begin{equation}
n^{HMT}(k) \propto 1/k^{4}
\end{equation}
and
\begin{equation}
\int_{k_{F}}^{2k_{F}}n^{HMT}(k)k^{2}dk \bigg/ \int_{0}^{2k_{F}}n(k)k^{2}dk \simeq 20\%
\end{equation}
and keeping
\begin{equation}
 n^{HMT}_{p}(k)/n^{HMT}_{n}(k) \simeq \rho_{n}/\rho_{p}.
\end{equation}
Where $\rho_{n}$ and $\rho_{p}$ are, respectively, local neutron and proton densities.
With this nucleon momentum distribution, the nucleon average kinetic energy changes very small compared with that using the ideal Fermi-Gas model. I thus neglect this difference in heavy-ion collisions at intermediate energies. 

\begin{figure}[th]
\centering
\includegraphics[width=0.5\textwidth]{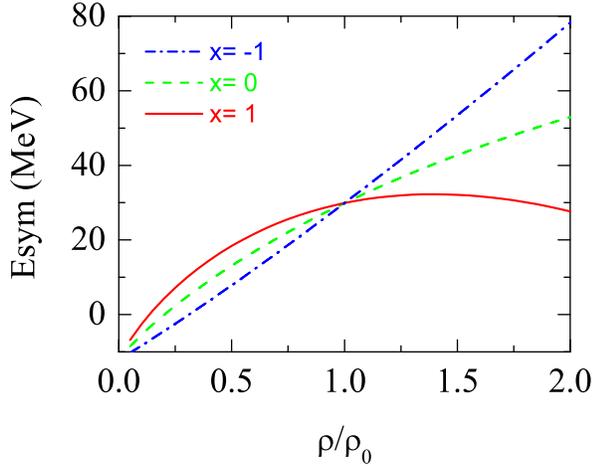}
\caption{The corresponding density-dependent symmetry energy with the single particle potential of Eq.~(\ref{buupotential}) with different $x$ parameters.} \label{esym}
\end{figure}
The isospin- and momentum-dependent single nucleon mean-field
potential is expressed as
\begin{eqnarray}
U(\rho,\delta,\vec{p},\tau)&=&A_u(x)\frac{\rho_{\tau'}}{\rho_0}+A_l(x)\frac{\rho_{\tau}}{\rho_0}\nonumber\\
& &+B(\frac{\rho}{\rho_0})^{\sigma}(1-x\delta^2)-8x\tau\frac{B}{\sigma+1}\frac{\rho^{\sigma-1}}{\rho_0^\sigma}\delta\rho_{\tau'}\nonumber\\
& &+\frac{2C_{\tau,\tau}}{\rho_0}\int
d^3\,\vec{p^{'}}\frac{f_\tau(\vec{r},\vec{p^{'}})}{1+(\vec{p}-\vec{p^{'}})^2/\Lambda^2}\nonumber\\
& &+\frac{2C_{\tau,\tau'}}{\rho_0}\int
d^3\,\vec{p^{'}}\frac{f_{\tau'}(\vec{r},\vec{p^{'}})}{1+(\vec{p}-\vec{p^{'}})^2/\Lambda^2},
\label{buupotential}
\end{eqnarray}
where $\rho_0$ denotes the saturation density, $\tau, \tau'$=1/2(-1/2) is for neutron potential $U_{n}$ (proton potential $U_{p}$),
$\delta=(\rho_n-\rho_p)/(\rho_n+\rho_p)$ is the isospin asymmetry,
and $\rho_n$, $\rho_p$ denote neutron and proton densities,
respectively. Considering the effect of nucleon-nucleon short-range correlations \cite{yong20151,yong20152}, the parameter values $A_u(x)$ = 33.037 - 125.34$x$
MeV, $A_l(x)$ = -166.963 + 125.34$x$ MeV, B = 141.96 MeV,
$C_{\tau,\tau}$ = 18.177 MeV, $C_{\tau,\tau'}$ = -178.365 MeV, $\sigma =
1.265$, and $\Lambda = 630.24$ MeV/c.
The saturation properties of symmetric and asymmetric
nuclear matter under the action of the modified momentum
distributions are kept, i.e.,
the saturation density $\rho_{0}$ = 0.16
fm$^{-3}$, the binding energy $E_{0}$ = -16 MeV, the
incompressibility $K_{0}$ = 230 MeV, the isoscalar effective mass
$m_{s}^{*} = 0.7 m$, the single-particle potential
$U^{0}_{\infty}$ = 75 MeV at infinitely large nucleon momentum at
saturation density in symmetric nuclear matter, the symmetry
energy $Esym(\rho_{0})$ = 30 MeV, the kinetic symmetry
energy $E_{sym}^{kin}(\rho_{0})$ = 0 MeV, the symmetry potential
$U^{sym}_{\infty}$ = -100 MeV at infinitely large nucleon momentum
at saturation density.
$f_{\tau}(\vec{r},\vec{p})$ is
the phase-space distribution function at coordinate $\vec{r}$ and
momentum $\vec{p}$ and solved by using the test-particle method
numerically. Different symmetry energy's stiffness parameters $x$
can be used in the above single nucleon potential to mimic
different forms of the symmetry energy predicted by various
many-body theories without changing any
property of the symmetric nuclear matter and the symmetry energy
at normal density. Fig.~\ref{esym} shows the corresponding symmetry energy with the single particle potential of Eq.~(\ref{buupotential}) with different $x$ parameters. $x = 1, 0, -1 $ cases respectively correspond to the slopes ($L (\rho_{0}) \equiv 3\rho_{0}dEsym(\rho)/d\rho$) of 37, 87, 138 MeV \cite{yong20152}.

For baryon resonance $\Delta$ potential, the forms of
\begin{eqnarray}
\begin{split}
U^{\Delta^-}&=U_{n},\\
U^{\Delta^0}&=\frac{2}{3}U_{n}+\frac{1}{3}U_{p},\\
U^{\Delta^+}&=\frac{1}{3}U_{n}+\frac{2}{3}U_{p},\\
U^{\Delta^{++}}&=U_{p}
\end{split}
\end{eqnarray}
are used.
The isospin-dependent baryon-baryon ($BB$) scattering cross section in medium $\sigma
_{BB}^{medium}$ is reduced compared with their free-space value
$\sigma _{BB}^{free}$ by a factor of
\begin{eqnarray}
R^{BB}_{medium}(\rho,\delta,\vec{p})&\equiv& \sigma
_{BB_{elastic, inelastic}}^{medium}/\sigma
_{BB_{elastic, inelastic}}^{free}\nonumber\\
&=&(\mu _{BB}^{\ast }/\mu _{BB})^{2},
\end{eqnarray}
where $\mu _{BB}$ and $\mu _{BB}^{\ast }$ are the reduced masses
of the colliding baryon pairs in free space and medium,
respectively. The effective mass of baryon in isospin asymmetric nuclear matter
is expressed as
\begin{equation}
\frac{m_{B}^{\ast }}{m_{B}}=1/(1+\frac{m_{B}}{p}\frac{%
dU}{dp}).
\end{equation}

In this semi-classical transport model IBUU, I
actually do not propagate the full spectral function of pions \cite{pionp1,pionp2,pionp3,pionp4,pionp5}.
I describe the particles as classical quasi-particles by adding an effective optical potential for the pions in the nuclear medium. A density- and momentum-dependent pion potential including isoscalar and isovector contributions is used \cite{pionp6,yong20153}. It is repulsive at low pionic momenta but attractive at high pionic momenta. The isoscalar potential is overall positive bu the isovector potential is positive for $\pi^{-}$ while negative for $\pi^{+}$.

The free elastic proton-proton cross section $\sigma_{pp}$ and neutron-proton cross section $\sigma_{np}$ are taken from experimental data.
The free elastic neutron-neutron cross section $\sigma_{nn}$ is assumed to be the same as the $\sigma_{pp}$ at the same center of mass energy. Other baryon-baryon free elastic cross sections are assumed to be the same as that of nucleon-nucleon elastic cross section at the same center of mass energy.
The nucleon-nucleon free inelastic isospin decomposition cross sections
\begin{equation}
\begin{split}
\sigma^{pp\rightarrow n\Delta^{++}}&=\sigma^{nn\rightarrow p\Delta^{-}}=\sigma_{10}+\frac{1}{2}\sigma_{11},\\
 \sigma^{pp\rightarrow p\Delta^{+}}&=\sigma^{nn\rightarrow n\Delta^{0}}=\frac{3}{2}\sigma_{11},\\
 \sigma^{np\rightarrow p\Delta^{0}}&=\sigma^{np\rightarrow n\Delta^{+}}=\frac{1}{2}\sigma_{11}+\frac{1}{4}\sigma_{10}
\end{split}
\end{equation}
are parameterized via
\begin{equation}
    \sigma_{II'}(\sqrt{s})=\frac{\pi(\hbar c )^{2}  }{2p^{2}}\alpha(\frac{p_{r}}{p_{0}})^{\beta}\frac{m_{0}^{2}\Gamma^{2}(q/q_{0})^{3}}
    {(s^{\ast}-m_{0}^{2})^{2}+m_{0}^{2}\Gamma^{2}}
\end{equation}
with $I$ and $I'$ being the initial state and final state isospins of two nucleons.
The parameters $\alpha, \beta, m_{0}, \Gamma$ as well as other kinematic quantities can be found in Ref.~ \cite{VerWest1982}. For neutron and proton, the isospin-dependent pauli-blockings with neighboring 64 lattices interpolation are used.

The mass of produced $\Delta$ in an inelastic
nucleon-nucleon collision is generated according to a modified Breit-Wigner function \cite{dan91}
\begin{equation}
    P(m_{\Delta})=\frac{p_{f}m_{\Delta}\times4m_{\Delta0}^{2}\Gamma_{\Delta}}
    {(m_{\Delta}^{2}-m_{\Delta0}^{2})^{2}+m_{\Delta0}^{2}\Gamma_{\Delta}^{2}}.
\end{equation}
Here $m_{\Delta0}$ is the centroid of the resonance and $\Gamma_{\Delta}$ is the width of the resonance and $p_{f}$ is the center of mass momentum in the $N \Delta$ channel.
The cross section for the two-body free inverse reaction is calculated by the modified detailed balance \cite{li1993}, which takes into account the
finite width of baryon resonance \cite{dan91}
\begin{equation}
    \sigma_{N\Delta\rightarrow NN}=\frac{m_{\Delta}p_{f}^{2}\sigma_{NN\rightarrow N\Delta}}{2(1+\delta)p_{i}}
    \bigg/\int_{m_{\pi}+m_{N}}^{\sqrt{s}-m_{N}}\frac{dm_{\Delta}}{2\pi}P(m_{\Delta}).
\end{equation}
The factor $(1+\delta)$ takes into account the case of having two identical nucleons in the final state. Here $p_{f}$ and $p_{i}$ are the nucleon center of mass momenta in the $NN$ and $N\Delta$ channels.

The baryon resonance $\Delta$ is assumed to be produced isotropically in the nucleon-nucleon center of mass,
and the decay of $\Delta\rightarrow \pi N$ also has an isotropic angular distribution in the
$\Delta$ rest frame.
The width of $\Delta$ resonance is
given by \cite{kita1986}
\begin{equation}
    \Gamma_\Delta=\frac{0.47q^{3}}{m_{\pi}^{2}[1+0.6(q/m_{\pi})^{2}]}.
\end{equation}
Where $q$ is the pion momentum in the $\Delta$ rest frame, which is related to
the mass of $\Delta$ resonance
\begin{equation}
   q= \sqrt{(\frac{m_{\Delta}^{2}-m_{n}^{2}+m_\pi^{2}}{2m_{\Delta}})^{2} -m_\pi^{2}                      }.
\end{equation}
The mass of $\Delta$ resonance in $\pi+N\rightarrow \Delta$ process is uniquely determined
by the reaction kinematics.
A Monte Carlo sampling of the $\Delta$ decay
is carried out according to the probability
\begin{equation}
    P_{deacy}=1-exp(-dt\Gamma_{\Delta}/\hbar).
\end{equation}

The Breit-Wigner form of the resonance formation in the pion-nucleon interaction is used \cite{li2001} \begin{equation}
    \sigma_{\pi+N}=\sigma_{max}(\frac{q_{0}}{q})^{2}\frac{\frac{1}{4}\Gamma_{\Delta}^{2}}{(m_{\Delta}-m_{\Delta 0})^{2}+\frac{1}{4}\Gamma_{\Delta}^{2}}.
\end{equation}
Where $q_{0}$ is the pion momentum at the centroid $m_{\Delta 0}$=1.232 GeV of the resonance mass distribution.
The maximum cross sections $\sigma_{max}$ are
\begin{eqnarray}
\begin{split}
\sigma_{max}^{\pi^{+}p\rightarrow \Delta^{++}}&= \sigma_{max}^{\pi^{-}n\rightarrow \Delta^{-}}=200 ~mb,\\
\sigma_{max}^{\pi^{-}p\rightarrow \Delta^{0}}&= \sigma_{max}^{\pi^{+}n\rightarrow \Delta^{+}}=66.67~mb,\\
\sigma_{max}^{\pi^{0}p\rightarrow \Delta^{+}}&= \sigma_{max}^{\pi^{0}n\rightarrow \Delta^{0}}=133.33~mb.
\end{split}
\end{eqnarray}

For the baryon-baryon inelastic cross section, the angle is determined
by assumed isotropy of scattering. For the baryon-baryon elastic scattering, the angular distribution is taken as \cite{cu1981}
\begin{equation}
    \frac{d\sigma_{el}}{d\Omega}\propto e^{bt}
\end{equation}
and $t$ = $-2p^{2}(1-\cos\theta)$ is the negative of the square of the momentum transfer in the center of mass, $p$ is the momentum of one particle in the center of mass and
\begin{equation}
    b=\frac{6[3.65(\sqrt{s}-1.8766)]^{6}}{1+[3.65(\sqrt{s}-1.8766)]^{6}}.
\end{equation}
The final momenta of particles can then be obtained via Energy-Momentum conservation \cite{bertsch}.

\section{Results and Discussions}

\begin{figure}[th]
\centering
\includegraphics[width=0.5\textwidth]{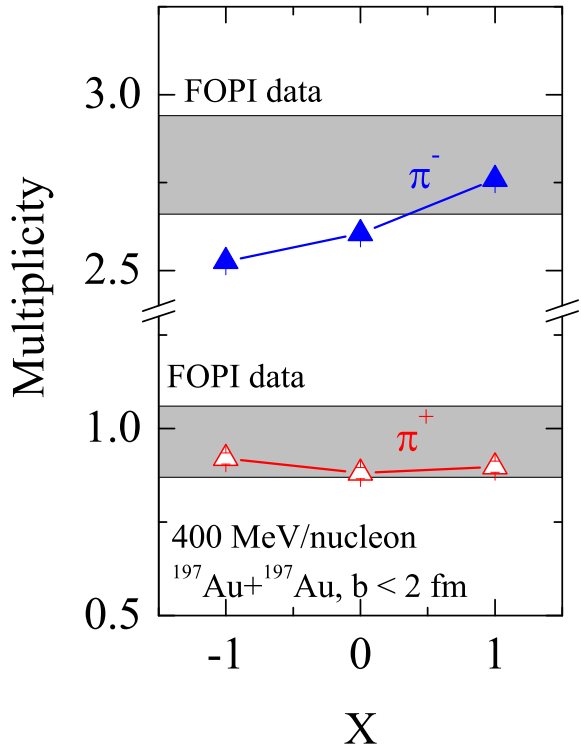}
\caption{Charged pion yields in Au+Au
reaction at 400 MeV/nucleon with different symmetry energies. The shadow region denotes the FOPI data \cite{Reisdorf10}.} \label{multipion}
\end{figure}
To check my transport model on pion production, it is instructive to see the production of charged pions in central Au + Au reaction at 400 MeV/nucleon beam energy since pion production data in this reaction are available. Fig.~\ref{multipion} shows the numbers of charged pions produced with different symmetry energies. It is first seen that both produced $\pi^{-}$ and $\pi^{+}$ by my IBUU model fit the FOPI experimental data quite well. Comparing produced $\pi^{-}$ and $\pi^{+}$ based on my IBUU model, it is seen that sensitivity of the number of produced $\pi^{-}$ to the symmetry energy is evidently larger than that of $\pi^{+}$. This is because the $\pi^{-}$'s are mainly from neutron-neutron collisions, thus more sensitive to the symmetry energy \cite{lyz05}. It is also seen that for the soft symmetry energy $x$ = 1, the produced $\pi^{-}$ fits the FOPI experimental data very well. With stiffer symmetry energies $x = 0, -1$, the model gives smaller $\pi^{-}$ numbers than experimental data.

\begin{figure}[th]
\centering
\includegraphics[width=0.5\textwidth]{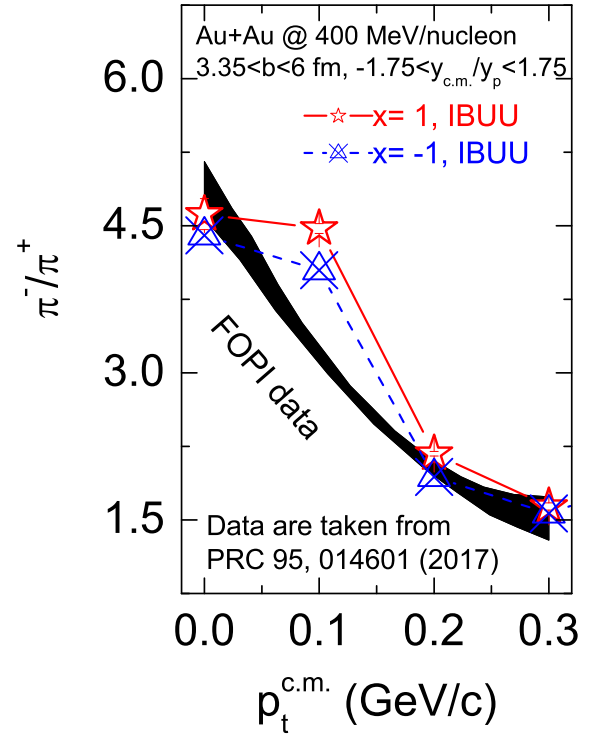}
\caption{Transverse momentum distribution of the $\pi^{-}/\pi^{+}$ ratio in Au+Au
reaction at 400 MeV/nucleon with different symmetry energies. The shadow region denotes the FOPI data \cite{cozma17}.} \label{ptpion}
\end{figure}
Fig.~\ref{ptpion} shows the transverse momentum distribution of the ratio of $\pi^{-}$ over $\pi^{+}$ yields in Au+Au reaction at 400 MeV/nucleon with different symmetry energies. It is seen that, with different symmetry energies, the ratio of $\pi^{-}$ and $\pi^{+}$ yields in Au+Au reaction at 400 MeV/nucleon given by my IBUU model fits the FOPI data very well, except for that around $p_{t}$ = 0.1 GeV/c. One can also see that the current transverse momentum distribution of FOPI pion data can not constrain the stiffness of the density-dependent symmetry energy. While my previous studies show that, in mid-central Au+Au reaction at 400 MeV/nucleon, in the direction perpendicular to the reaction plane, the $\pi^{-}/\pi^{+}$ ratio especially at high kinetic energies may exhibit significant sensitivity to the symmetry energy \cite{gao13}.

With this transport model, in the following, I try to give predictions on the single and double $\pi^{-}/\pi^{+}$ ratios as a function of kinetic energy in isotope reactions of $^{132}$Sn+$^{124}$Sn and $^{108}$Sn+$^{112}$Sn at 300 MeV/nucleon incident beam energy since related pion measurements are ongoing at Radioactive Isotope Beam Facility (RIBF) at  RIKEN in Japan \cite{exp1,exp2}.

\begin{figure}[th]
\centering
\includegraphics[width=0.5\textwidth]{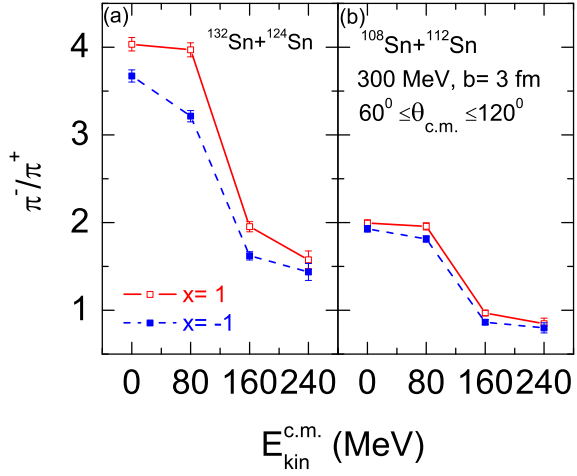}
\caption{The ratios of $\pi^{-}/\pi^{+}$ as a function of kinetic energy in isotope reaction systems of $^{132}$Sn+$^{124}$Sn and $^{108}$Sn+$^{112}$Sn at 300 MeV/nucleon incident beam energy with stiff (x= -1) and soft (x= 1) symmetry energies. $\theta_{cm}$ is polar angle relative to the incident beam direction.} \label{single}
\end{figure}
\begin{figure}[th]
\centering
\includegraphics[width=0.5\textwidth]{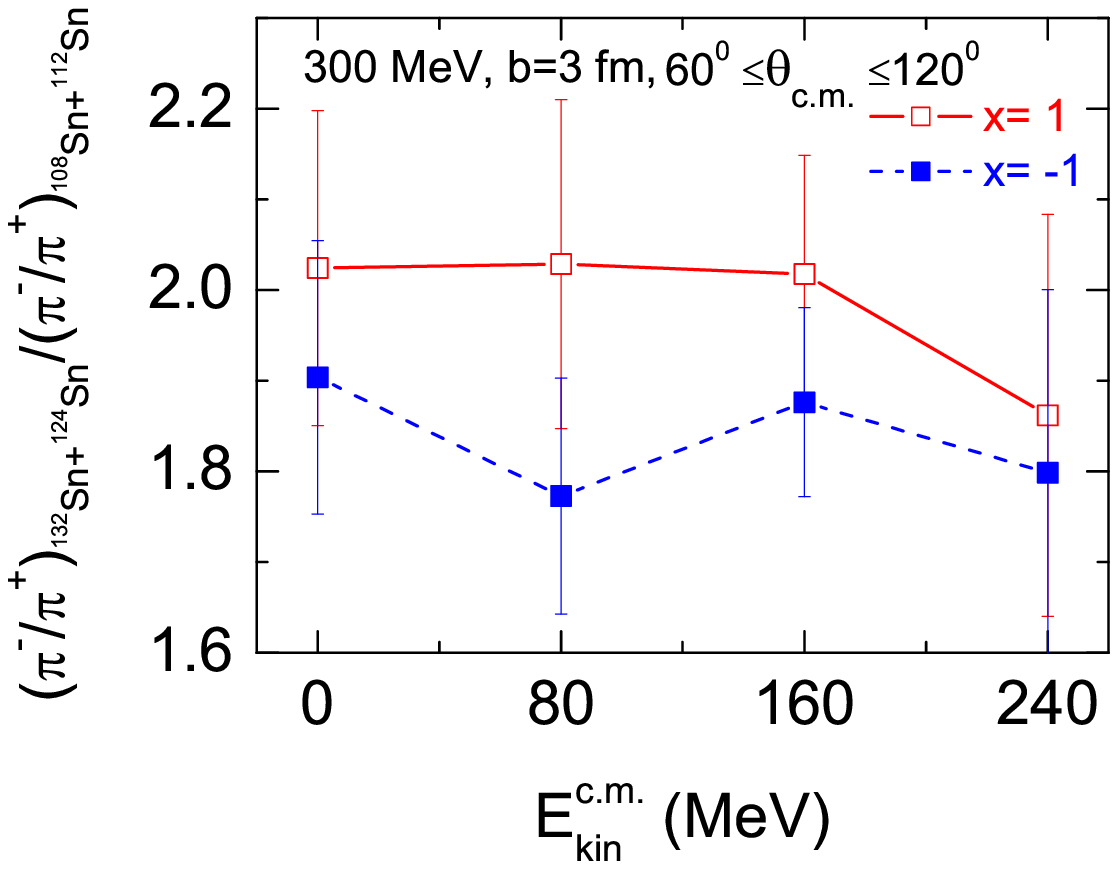}
\caption{The double ratio of $\pi^{-}/\pi^{+}$ as a function of kinetic energy in isotope reaction systems of $^{132}$Sn+$^{124}$Sn over $^{108}$Sn+$^{112}$Sn at 300 MeV/nucleon incident beam energy with stiff (x= -1) and soft (x= 1) symmetry energies. } \label{double}
\end{figure}
\begin{figure}[th]
\centering
\includegraphics[width=0.5\textwidth]{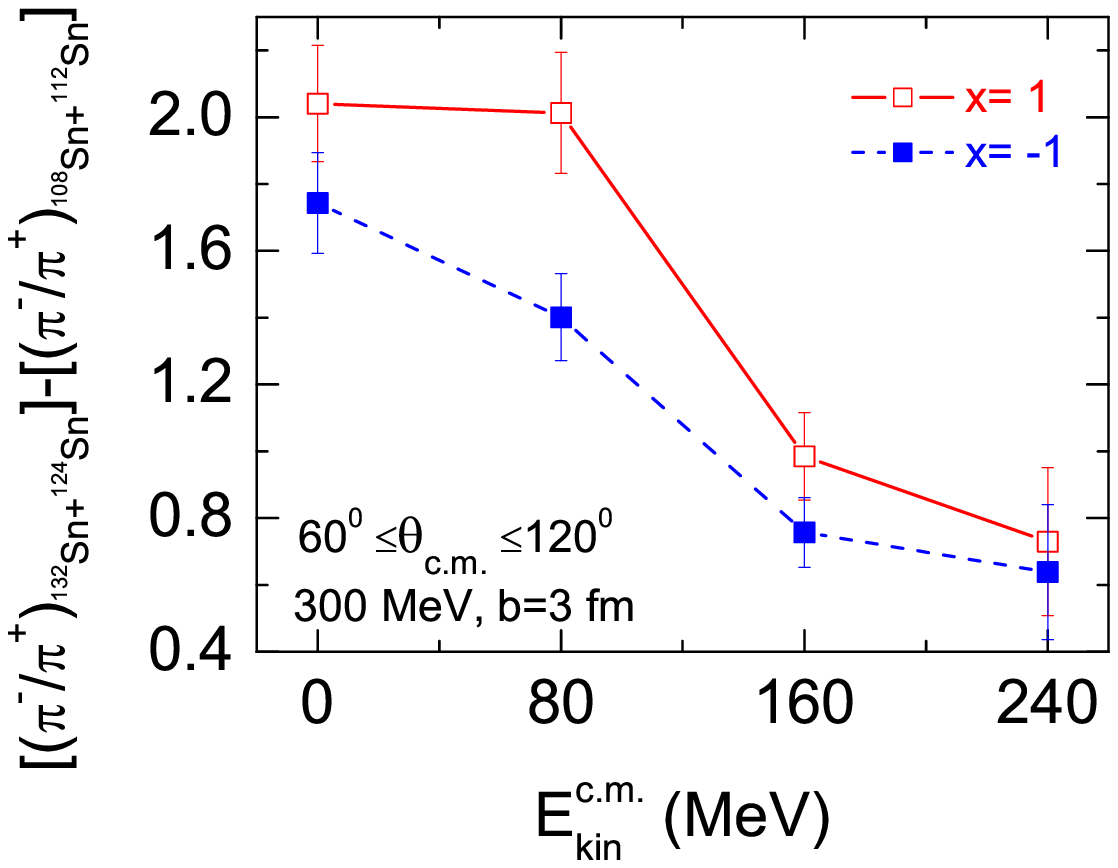}
\caption{The subtracted ratio of $\pi^{-}/\pi^{+}$ as a function of kinetic energy in isotope reaction systems of $^{132}$Sn+$^{124}$Sn and $^{108}$Sn+$^{112}$Sn at 300 MeV/nucleon incident beam energy with stiff (x= -1) and soft (x= 1) symmetry energies. } \label{sb}
\end{figure}
\begin{figure}[th]
\centering
\includegraphics[width=0.5\textwidth]{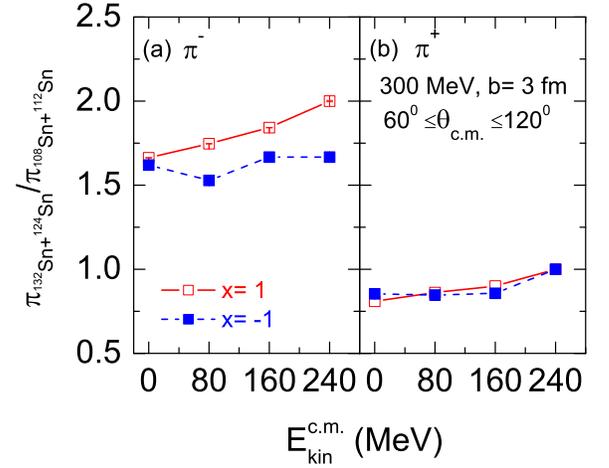}
\caption{Isoscaling ratios of $\pi^{-}$ (left panels) and $\pi^{+}$ (right panels)
from central collisions of $^{132}$Sn+$^{124}$Sn and $^{108}$Sn+$^{112}$Sn at 300 MeV/nucleon incident beam energy with stiff (x= -1) and soft (x= 1) symmetry energies.
} \label{isopi}
\end{figure}
Fig.~\ref{single} shows the single ratios of the $\pi^{-}/\pi^{+}$ in neutron-rich and neutron-deficient reaction systems $^{132}$Sn+$^{124}$Sn and $^{108}$Sn+$^{112}$Sn at a beam energy of 300 MeV/nucleon. Owing to more neutron-neutron collisions, it is clearly seen that the ratio of $\pi^{-}/\pi^{+}$ is higher in neutron-rich reaction system than that in neutron-deficient reaction system. And due to larger asymmetry in neutron-rich reaction system, the effects of the symmetry energy on the $\pi^{-}/\pi^{+}$ ratio are evidently larger than that in neutron-deficient reaction system.

To reduce the Coulomb and some other isospin-independent systematic errors, it is better to make the double ratio of $\pi^{-}/\pi^{+}$ in neutron-rich and neutron-deficient reaction systems, i.e., at different kinetic energy points, one makes the ratio of the $\pi^{-}/\pi^{+}$ ratio from the neutron-rich reaction system over that from the neutron-deficient reaction system \cite{yong2006}. Fig.~\ref{double} shows the double ratio of $\pi^{-}/\pi^{+}$ from the neutron-rich and neutron-deficient isotope Sn reaction systems. It is seen that the trend of the double $\pi^{-}/\pi^{+}$ ratio as a function of kinetic energy becomes flat compared with the trends of single $\pi^{-}/\pi^{+}$ ratios as a function of kinetic energy shown in Fig.~\ref{single}. The other method to reduce the systematic errors is the subtracted ratio of $\pi^{-}/\pi^{+}$ \cite{tsang2017}. Fig.~\ref{sb} shows the subtracted ratio of $\pi^{-}/\pi^{+}$. It is seen that the present IBUU gives a descending trend of the subtracted ratio of $\pi^{-}/\pi^{+}$ as a function of kinetic energy. Because detecting negative pion and
constructing the $\pi^{-}$ isoscaling ratio are much easier than
that of $\pi^{+}$ \cite{tsang2017}, I also plot Fig.~\ref{isopi}, the isoscaling ratios of $\pi^{-}$ and $\pi^{+}$. It is seen that the isoscaling ratio of $\pi^{-}$ is more sensitive to the symmetry energy than the isoscaling ratio of $\pi^{+}$.

\section{Conclusions}

By considering the effects of nucleon-nucleon short-range correlations and in-medium reduced baryon-baryon cross sections as well as pion mean-field potential in the isospin-dependent IBUU transport model, I successfully reproduced pion production in Au+Au reactions at 400 MeV/nucleon. To obtain the information on the density-dependent symmetry energy, predictions on single and double pion ratios in isotope Sn reactions in $^{132}\rm {Sn}+^{124}\rm {Sn}$ and $^{108}\rm {Sn}+^{112}\rm {Sn}$ at 300 MeV/nucleon are made for experiments at RIKEN/Japan. These studies may help one to model pion production in heavy-ion collisions at intermediate energies and to constrain the density-dependent symmetry energy by pion production using a wide variety of advanced new facilities \cite{yonggwz}, such as the Facility for Rare Isotope Beams (FRIB) in
the US, the Facility for Antiproton and
Ion Research (FAIR) at GSI in Germany, the Radioactive Isotope Beam Facility (RIBF)
at RIKEN in Japan, the Cooling Storage Ring on the Heavy Ion Research Facility at IMP (HIRFL-CSR) in China \cite{xiao17}, the Korea Rare Isotope Accelerator (KoRIA) in Korea.

\section{Acknowledgments}

The work is supported by the National Natural Science
Foundation of China under Grant Nos. 11375239, 11775275, 11435014.

\end{document}